\documentstyle[epsfig, proceedings, namedreferences]{crckapb} 
\begin{opening}
\title{Circumstellar Disks}
\author{Steven V. W. Beckwith}
\institute{Max-Planck-Institut f\"ur Astronomie\\
\& Space Telescope Science Institute}
\end{opening}
\begin{document}

\section{Introduction}
Circumstellar disks usually contain only a few percent of the total material going into a young star after the main collapse has stopped and the surrounding molecular cloud is cleared away.  Yet the disks are of great interest to the study of star formation, perhaps as great as the stars themselves, because the disks may build planetary systems.   The planet Earth contains less than one millionth of the mass of the Sun, but it is probably the most interesting body in the Solar System, certainly to us.  Beckwith \& Sargent (1996) argue that the currently known properties of disks are evidence that other planetary systems are common in the Galaxy and discuss the reasons for the interest in disk properties; that article provides a broad introduction to the subject.

The purpose of this chapter is to provide a tutorial on how observations of the radiation from disks may be used to elicit their physical characteristics.  In keeping with the spirit of the Crete meeting, the treatment is not a comprehensive review nor will it give a complete analysis of each method used to tease the disk properties from faint light observed with telescopes.  Rather, the idea is to show that basic intuition about disk physics is easily related to what is observed and provide a general foundation for the understanding of more elaborate theoretical calculations.  

Early disk models assumed that matter is confined to a very thin plane extending from the stellar surface to a sharp outer edge more than 100 AU from the star.  The disk energy balance was attributed to accretion of matter through the disk.  This oversimplified picture has been modified by a careful treatment of the underlying physics, and the more modern view is that the disk flares gently, often with an inner edge at some distance from the star, and is heated mainly by radiation as opposed to accretion.  Some disks are surrounded by spheroidal ``halos'' that trap radiation and contain strong outflows.  Most of the young disks are accompanied by mass loss in columns along the polar axes that contribute to the total energy budget.  Although it is not always possible to derive disk characteristics unambiguously from observations, most of the intuitive interpretations have been supported by increasingly better data and improved angular resolution images allowing us to separate the different components of a star/disk system directly.

The article is organized along the following questions: 
\begin{enumerate}
\item What are the expected disk properties based on the theory of Solar System formation?
\item How do we identify disks?
\item How do we determine physical properties of disks from radiation?
\item Do the observed properties show that disks are interesting?
\end{enumerate}
Because this article is a tutorial, some of the material is adopted from articles that I co-authored for Nature (Beckwith \& Sargent 1996) and Protostars and Planets IV (Beckwith, Henning, and Nakagawa 1999).

\section{The early Solar System}
It is generally agreed that a flat layer of gas and dust - a disk - orbited the early Sun and provided the material which later made up the Earth, Mars, Jupiter, and the other planets (Safronov 1969; Wood \& Morfill 1988; Cameron 1988). The young Sun and the circumsolar disk were born from an extended cloud of gas and dust that was assembled from the detritus of dying stars and remnants of the early universe that collapsed under its own gravity.  The material accumulated quickly onto the central proto-Sun but with enough residual angular momentum to prevent some from spiraling inwards - the exact proportion remaining in orbit is not known but should have been a considerable fraction of the total mass (Shu, Adams, \& Lizano 1987; Bodenheimer 1995).  The average angular momentum of the collapsing region defined a rotation axis around which the orbits quickly stabilized, creating a disk with a thickness much smaller than its radius, at least within the regions now containing the giant planets.  The formation of the stable disk probably occurred over about $10^5$ years after the onset of free fall collapse (Shu et~al. 1993), almost instantaneously in cosmic time.

As the central proto-Sun evolved, the solid particles in the orbits settled to a dense layer in the mid-plane of the disk and began to stick together as they collided (Safronov 1969; Weidenschilling 1987; Mizuno et~al. 1988).  During the next $10^4$ to $10^5$ years, large rocks and small asteroids grew gradually from the small dust particles (Weidenschilling \& Cuzzi 1993).  When the gravitational pull of the largest asteroids was sufficient to attract neighboring pebbles and rocks, they grew even more rapidly to the size of small planets (Wetherill \& Stuart 1993); gravity was important for bodies more than 10\,km across.  The terrestrial planets are large accumulations of solid particles that grew from the collisions between these smaller bodies. In the outer parts of the disk, a few such solid cores became large enough (10 Earth masses) to accrete gas (Mizuno 1980; Stevenson 1982), the dominant reservoir of mass, and gave rise to the giant gas planets (Wetherill 1990).  Temperatures close to the proto-Sun were presumably too high to allow gas accretion.  The planet building phase is thought to have taken between $\sim 10^7$ and a few times $10^8$ years, although the cores probably developed quickly, within the first $10^6$ years or so. These timescales are not very well constrained by data, and it may well be observations of developing planetary systems around other stars that tell us how planets are really built.

By analogy, we expect circumstellar disks to contain a few percent or more of the stellar mass, to extend at least 50\,AU from the central star, to be relatively flat, and to be free from disruption for at least a few million years if they are to create the rocky cores needed to build the large planets.  These characteristics certainly represent only a subset of the disks that accompany star formation.  Conceivably, a disk with substantially lower mass ($10^{-6}$\,M$_\odot$), size (a few AU), and lifetime ($\sim 10^6$\,yr) could create terrestrial-like planets suitable for life without the presence of gas giants.  In principle, even larger, more massive disks could accompany the birth of stars much more massive than the Sun.  Although we must keep an open mind about the characteristics the constitute a disk, the early Solar disk provides a framework to identify those disks that may become interesting for planet formation.  

\section{How do we know that disks exist?}
Soon after the discovery that T Tauri stars -- very young stars of approximately solar mass -- had more radiation at infrared wavelengths than the photospheres should emit, Lynden-Bell and Pringle (1974) suggested that most of their peculiar characteristics might be explained by circumstellar disks.  Their suggestion was based on the unusual spectral energy distributions: the stars radiated too much ultraviolet light {\it and} too much infrared light at the same time.  A disk could account for the ultraviolet light through emission from the {\it boundary layer} between the star and the inner edge of the disk, in which matter from the disk suddenly accreted onto the star, slowing down from Keplerian speeds to essentially zero speed so rapidly that the radiation temperatures are tens of thousands of Kelvin.  The infrared light was radiation from the outer parts of the disk resulting from energy liberated as the matter slowly spiraled to smaller radii eventually to accrete through the boundary layer.  One of their strong predictions was that the long wavelength infrared radiation would follow a power law, $F_\nu \propto \nu^{1 \over 3}$, where $F_\nu$ is the flux density, and $\nu$ is the frequency of the radiation.  This result for the release of accretion energy through a disk is quite general.

Even with no accretion, a disk will be heated by radiation from the star itself.  The dust grains in the disk absorb stellar radiation and re-radiate in the infrared to maintain thermal balance.  Remarkably, the spectral energy distribution also follows a power law with the same exponent as for accretion over a broad range of wavelengths: $F_\nu \propto \nu^{1 \over 3}$ between about 5 and 100\,$\mu$m depending on the luminosity of the star. At wavelengths short ward of about 3\,$\mu$m, the flux density stops increasing due to the inner radius of the disk, and at long wavelengths the power law becomes steeper due to the outer edge of the disks.  The most complete treatment for a flat disk is given by Adams, Lada, and Shu (1988).

When Lynden-Bell and Pringle made their suggestion, the long wavelength SEDs of disks could only be measured to about 10\,$\mu$m.  The SEDs could easily be explained over this limited spectral range by other distributions of dust near the stars.  Although prescient, the predictions of the early disk theory went untested for nearly a decade.

A few years after the SED calculations, Els\"asser and Staude (1978) discovered that several young stars had rather high degrees of linear polarization in their optical light.  They explained the polarization as scattered light from dust grains arranged symmetrically above and below a star that was obscured by a planar or toroidal distribution of dust oriented perpendicular to the line of sight and parallel to the direction of the polarization.  These observations suggested that the dust distribution around the stars was axisymmetric and flattened relative to a spherical halo.

The most striking demonstrations of axisymmetry in T Tauri stars are the well collimated jets seen in images of ionized lines.  Mundt and Fried (1983) discovered the first of many young stellar jets in their image of HL~Tau. The jets implied a strong axisymmetry near the stars, one component of which may be flattened disks perpendicular to the axis of the jets.  In the same year, Cohen (1983) used the Kuiper Airborne observatory to measured the first far infrared SED of HL~Tau, and he inferred the presence of a thin disk following the reasoning of Lynden-Bell and Pringle (1974).  One year later, two groups observed HL~Tau at angular resolutions several times better than previously possible and discovered elongated infrared emission reminiscent of disks (Beckwith et al. 1984; Grasdalen et al. 1984).  Following the evidence accumulated for some time on this object, both groups suggested that the disk might be the progenitor of a planetary system similar to the early solar nebula.  Subsequently, Beckwith et al. (1986) and Sargent and Beckwith (1987) made interferometric maps of the CO and $^{13}$CO emission and showed that the gas, too, was elongated as if in a disk; the direction of elongation was perpendicular to the jet and aligned with the major axis of the near infrared emission.  The latter paper also observed the velocity field and interpreted it as gas orbiting the star.  At this time, the evidence for a disk surrounding HL~Tau appeared to be quite strong.

These observations implied the distribution of material was disk-like but did not determine how much material was in a disk.  The interferometric maps showed that HL~Tau had continuum emission at 2.7\,mm that appeared to be thermal emission from dust particles.  With conservative assumptions about the radiative properties of the dust, the authors derived a mass for the HL~Tau disk of $\sim 0.1$\,M$_\odot$ with a considerable uncertainty. This mass is well above the minimum mass for the disk around the primitive solar nebula, 0.01\,M$_\odot$, thereby strengthening the case for a protoplanetary disk surrounding HL~Tau.  

At about the same time, astronomers analyzing the far infrared radiation from disks from the IRAS survey recognized that many T Tauri stars had strong far infrared excesses indicative of disks along the lines first noted by Cohen (1983).  Rucinski (1995), Beall (1987) and Adams, Lada, and Shu (1987) recognized that the IRAS observations were most easily interpreted as disk emission, although there was some controversy about the applicability of Lynden-Bell and Pringle's flat, black disk calculations. Adams, Lada, and Shu (1988) showed that a class of these sources, termed ``flat-spectrum sources'' (see the next section), required disks that were quite a bit warmer in their outer regions than predicted by the flat, black disk model.

Two major advances followed in 1989 and 1990.  Working with data from the IRAS sky survey, Cohen, Emerson, and Beichman (1989) and Strom et al. (1989) discovered that almost half the T Tauri stars in the Taurus-Auriga dark clouds had far infrared SEDs characteristic of heated dust, and the relatively low, visual extinction implied that the dust was in flattened distributions -- most likely disks.  This result meant that disks were common around young stars; HL~Tau was not an anamoly.  In the following year, Beckwith et al. (1990) found that a similar fraction of stars had emission at 1.3\,mm, from which they placed an even stronger limit on the flattening, essentially showing that disks were the only viable explanations for the data.  For the first time, they derived the distribution of masses for the disks around stars.  Not only were half the T Tauri stars surrounded by disks, the great majority of the disks had sufficient mass to build planetary systems like our own.  

\begin{figure}
  [!h]
  \begin{center}
    \leavevmode
  \centerline{\epsfig{file=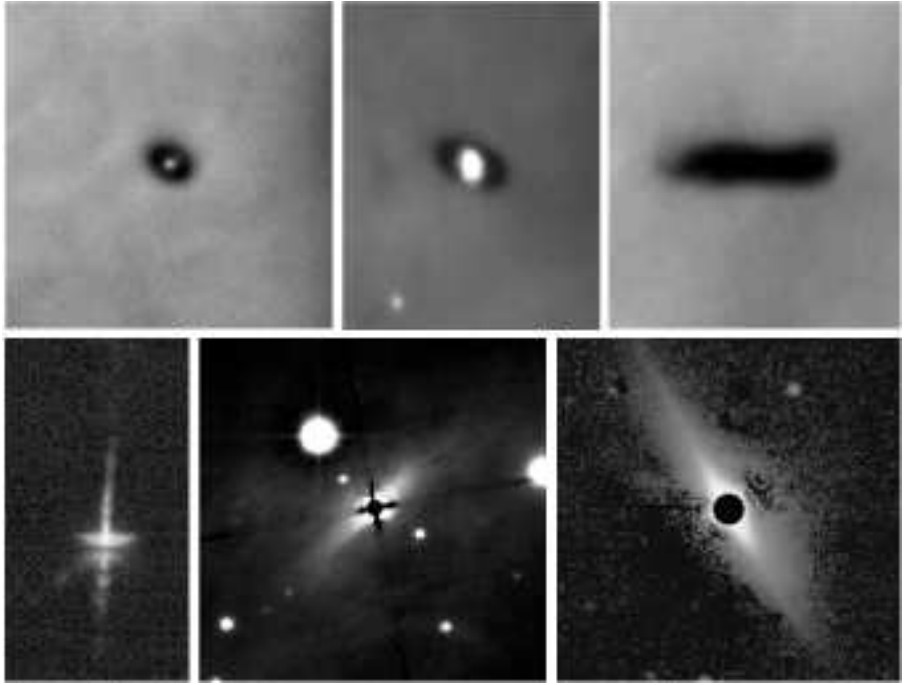}}
  \end{center}
  \vskip.1in
  \caption{\em These images of disks were observed with the Hubble Space Telescope.  The upper three images are disks seen in silhouette against the Orion nebula (McCaughrean \& O'Dell 1996).  The bottom three images show disks close to edge on.  The image of HH~30 (Burrows et~al. 1996) shows flaring and illumination of cone-shaped cavities in the surrounding halo due to residual cloud material.  The two images on the lower right are of main sequence disks: BD+31~643 (Kalas \& Jewitt 1997) and $\beta$~Pictoris (Kalas \& Jewitt 1995, and references therein).}
\label{DiskImages}
\end{figure}

Most specialists in star formation had by then accepted disks as the best explanation for the accumulated data on T Tauri stars.  But the evidence was largely indirect, and it was often possible to interpret the data sets with different theories for the distribution of dust near these stars.  The first pictures of disks taken with the Hubble Space Telescope demonstrated clearly that the dust distributions followed the theoretical pattern of a disk.  The image of HH~30 by Burrows et~al. (1996) shows a jet perpendicular to a dark disk seen edge-on with a gently flared shape, as expected for thermally supported disks (Kenyon \& Hartmann 1987).  Striking examples of disks are seen in silhouette against the Orion Nebula by O'Dell \& Wen (1994) and McCaughrean \& O'Dell (1996).  Several more images of disks like HH~30 have appeared recently (Stapelfeldt 1998; Padgett et~al. 1999).  Figure~\ref{DiskImages} shows several images of disks taken with the Hubble Space Telescope.

The images leave no doubt about the distribution of matter: the particles surrounding T Tauri stars are confined to flattened disks.  Many other observations -- very high resolution spectra (Hartmann et al. 1986), interferometric maps of the gas (Dutrey, Guilloteau, \& Simon 1994; Koerner \& Sargent 1995; Saito et~al. 1995) and polarization maps (Bastien \& Menard 1990; Whitney \& Hartmann 1992; Piirola, Scaltriti, \& Coyne 1992) -- are easily interpreted if disks are present and difficult to explain with other distributions of the matter.  Disks exist.

Compounded, the evidence for disks is overwhelming, but that does not mean that disks are present around every star or that individual stars may not have other distributions of matter.  For those stars that do have disks, we turn our attention to their properties and how we observe them.

\section{Spectral Energy Distributions}
Disks contain a mixture of gas and dust. Early in a disk's evolution -- the first few million years, say -- there are enough small particles to absorb and reradiate light from the star.  Dust warmed by the starlight radiates as a blackbody modified by the emissivity of the grains, which for small particles is approximately proportional to the frequency throughout the infrared region. The dust temperatures depend on the distance of a grain from the star.  Because a continuous disk contains particles at a wide range of distances from close to the star's surface to several hundred AU or more, dust temperatures span several orders of magnitude.  The result is a mixture of thermal emission across a broad spectrum from about 1\,$\mu$m to more than 100\,$\mu$m. 

The dust may be heated radiatively -- by the central stars, for example -- or by the gas via the loss of gravitational energy as it spirals through the disk to the star, called accretion heating.  Accretion heating might be important in the earliest stages of disk formation and could dominate the total energy budget for a disk within the innermost 10 stellar radii or so, but it is probably unimportant for the infrared spectral energy distributions discussed in this section (Chiang \& Goldreich 1996).  For this reason, we will examine only radiative heating mechanisms for disks in this article.

The study of accretion onto the star is a large topic in itself and well treated by other authors.  The interested reader is referred to Bertout (1989).  It has been suggested that disks may be heated by wave-driven accretion (Shu et~al. 1990), where the waves are first moment modes ($m=1$). This hypothesis has yet to find favor in the community, because it adds a complication to the physics for which there is no observational evidence, and radiative heating appears to be adequate to explain almost all observations.  The subject is not closed, however, and it may be that dynamical heating is seen to be important when the angular resolution of telescopes becomes good enough to map the density distributions of disks.

Young objects display a variety of different spectral energy distributions depending on their state of evolution.  At the very earliest stages, the SEDs are dominated not by the disks but by envelopes surrounding a central luminosity source most likely in free fall onto the star/disk system.  After some time, these envelopes dissipate, either because all the matter is accreted onto the star or disk or because winds from the central objects blow away the remaining material.  Lada (1987) invented a classification scheme to describe the sequence of SEDs that result from this process, a sequence generally used today to classify ages on the basis of SEDs.  In the language of this scheme, we are concerned with Class II sources, sources in which the disk dominates the SEDs, and the envelope emission is negligible.  The younger objects are treated by C. J. Lada in his chapter.  However, Chiang \& Goldreich (1998) show that some Class I sources can be explained as standard T~Tauri star disks viewed at high inclination, and Padgett et~al. (1999) recently published infrared images of Class I sources that show almost all of them to be edge-on disks, so we must reserve judgment about the interpretation of individual objects based solely on their spectral class.

\subsection{Flat, black disks}
\begin{figure}
 [!h]
  \begin{center}
    \leavevmode
  \centerline{\epsfig{file=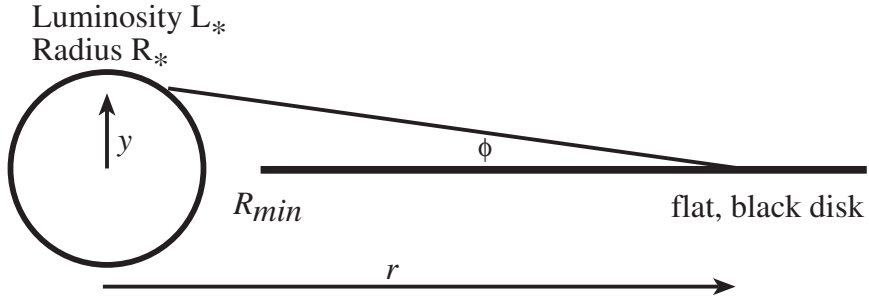,width=12cm}}
  \end{center}
  \vskip.1in
\caption{This sketch shows a star surrounded by a thin disk viewed in the plane of the disk.  It extends from an inner radius, $R_{min}$, to outer radius, $R_{max}$ (not shown), illuminated by a star of luminosity, $L_*$.  The disk is axisymmetric in all physical parameters.  A point at radius, $r$, in the disk is illuminated by a point on the star as shown with the variables labeled. The variable, $x$, is into the plane of the paper.}
\label{FlatDisk}
\end{figure}

To understand the infrared spectral energy distribution of a disk, it is useful to calculate the simplest case, a thin, black disk extending from the star to a distant boundary, specified by the outer radius, $R_{max}$.  Figure~\ref{FlatDisk} is a schematic diagram of the star/disk system viewed exactly in the plane of the disk and showing the variables used for the following analysis. 

The flux from the star that illuminates a point at radius, $r$, in the disk is given by integrating over the stellar surface as seen from the disk.  It is sufficient to consider only one surface (side) of the disk.  If $r$ is sufficiently large, so that the stellar diameter subtends a small angle, the flux is approximately:

\begin{eqnarray}
F_{illum} & \approx & {1\over r^2} \int_0^{R_*} 2 \sin\phi \int_0^{\sqrt{ R_*^2 - y^2 }} \sigma T_*^4\, dx\, dy, \label{FirstIllum} \\
          & \approx & {2\over r^2} \, \sigma T_*^4 \int_0^{R_*} \left( y \over r\right) \int_0^{\sqrt{ R_*^2 - y^2 }} dx\, dy, \\
          & \approx & {2\over 3} \left(R_* \over r\right)^3 \sigma T_*^4,
\label{Illum}
\end{eqnarray}
where $T_*$ is the effective temperature of the star.  The factor of $\sin\phi$ accounts for the illumination of the flat disk by a point on the star at distance, $y$, above the disk plane at distance, $r$, which is $\approx \left(y \over r\right)$ for large enough $r$.  This flux is totally absorbed by a black disk.

The disk radiates thermally as a blackbody, so:
\begin{equation}
F_{rad} = \sigma T^4(r),
\end{equation}
from which we derive,
\begin{equation}
T(r) \approx \left(2\over 3\right)^{1\over 4} T_* \left(r \over R_*\right)^{-{3\over 4}}.
\label{Teqn} 
\end{equation}
This result is nice, because it is easy to calculate the spectral energy distribution with free parameters that are, in principle, observable.  Assuming the star and disk are at distance, $D$, and oriented at angle, $\theta$, to the line of sight ($\theta = 0$ corresponds to face-on viewing), the flux density is:
\begin{eqnarray}
F_\nu & = & {1\over D^2} \int_{R_{min}}^{R_{max}} \cos\theta\, B_\nu\left[T(r)\right] 2\pi r\,dr \label{FnuInteg} \\
      & = & {1\over D^2}\, {4\pi h \nu^3 \over c^2} \left(k T_0 \over h \nu\right)^{8\over 3} R_{min}^2 \cos\theta\,\int_{x_{min}}^{x_{max}} {x dx \over exp(x^{3\over 4}) - 1}, \label{FlatInteg}
\end{eqnarray}
where we define $T_0 \equiv T(R_{min})$, the other physical constants are labeled conventionally, and we substituted variables to create the definite integral.  It is easy to evaluate (\ref{FlatInteg}) for any choice of inner and outer radii, but the beauty of this development is that for a wide range of wavelengths, the limits are essentially 0 and $\infty$, and the integral is a number.  To see this result, note that the limits are:
\begin{eqnarray}
x_{min} & \equiv & \left( h \nu \over k T(R_{min})\right)^{4\over 3} \\
x_{max} & \equiv & \left( h \nu \over k T(R_{max})\right)^{4\over 3} 
\end{eqnarray}
For most wavelengths, the emission from the inner and outer parts of the disk is negligible; $x_{min} \approx 0$ and $x_{max} \approx \infty$ for all practical purposes.  The definite integral is then:
\begin{equation}
\int_0^\infty {x dx \over exp(x^{3\over 4}) - 1} \approx 2.576
\end{equation}
and the spectral energy distribution becomes:
\begin{eqnarray}
F_\nu & \approx & {2.576\over D^2} \cos\theta\, {4\pi h \nu^3 \over c^2} \left(k T_0 \over h \nu\right)^{8\over 3} R_{min}^2 \\
    & \approx & C_1 D^{-2} \cos\theta\, T_0^{8\over 3} R_{min}^2 \nu^{1\over 3} \label{FlatSED}
\end{eqnarray}
Equation (\ref{Teqn}) shows how to determine $T(R_{min})$ for any value of $R_{min}$, if the stellar luminosity and source distance are known, and the disk orientation, $\theta$, is observed with high enough angular resolution.  The spectral energy distribution can be determined exactly.  This predicted SED is the standard by which most SEDs have been measured to determine if a star is surrounded by a disk.  

The approximations made to arrive at (\ref{FlatSED}) break down for high frequencies -- short wavelengths -- where the emission comes from near the inner disk boundary, $R_{min}$, and at low frequencies where the emission comes from near the outer disk boundary, $R_{max}$.  There is a further modification that becomes important when the inner disk boundary is close to the stellar radius, $R_*$.  In this case, the star and disk mutually heat one another to increase the temperatures of both objects slightly -- a kind of greenhouse effect -- and the actual SED is more complicated: Adams, Lada, and Shu (1988) calculate this effect exactly for the interested reader.  For our purposes, the approximations make very little error, since, as we will see, there are almost {\it no} examples of disks whose SEDs are fitted by this simple model.

We derived (\ref{FlatSED}) assuming a specific temperature distribution for the disk given by radiative heating.  We can generalize the result for arbitrary power law temperature distributions for disks that have other ways of establishing temperature distributions.  Assume that $T(r) = T_0 (r/R_{min})^{-q}$, then (\ref{FlatInteg}) becomes:
\begin{eqnarray}
F_\nu & \approx & {1\over D^2} {4\pi h \nu^3 \over c^2} \left(k T_0 \over h \nu\right)^{2\over q} R_{min}^2 \cos\theta\,\int_{x_{min}}^{x_{max}} {x dx \over exp(x^q) - 1}, \label{TempInteg} \\
      & \approx & C_2 D^{-2} \cos\theta\,T_0^{2\over q}R_{min}^2 \nu^{3-{2\over q}}, \label{TSED}
\end{eqnarray}
where we used the approximation that $x_{min} \approx 0$ and $x_{max} \approx \infty$.  The constant, $C_2$ includes the definite integral evaluated numerically for values of $q$ that allow convergence.

Equation~(\ref{TSED}) is useful for SEDs that are power law in frequency.  If the spectral index is defined as, $F_\nu \propto \nu^\alpha$, then (\ref{TSED}) tells us that $\alpha = 3 - {2 \over q}$.  The spectral index, $\alpha$ is observed directly.  Thus, it is always possible to derive the disk temperature distribution from the observed SED.  This result generally applies to black disks, and they need not be very flat to produce this result.  Even the absolute temperature at any radius may be derived if $D$ and $\theta$ are known.  It is, therefore, possible to estimate disk temperatures with only modest uncertainties from the observed SED, assuming only the disk contributes (see Beckwith et~al. 1990).

The flat disk model is quite useful for developing intuition to make more complicated models easier to understand.  At a wavelength, $\lambda$, the thermal emission comes from a limited range of radii in the disk.  The way to see this is to note that the integrand in (\ref{FlatInteg}) is maximum when $x \sim 0.5$, meaning $T(r) \approx 1.7 {h c / k \lambda}$, or $r(\lambda) \approx \left(0.6 k T(R_{min}) \lambda / h c \right)^{4\over 3} R_{min} \approx \left(\lambda / \lambda(R_{min}) \right)^{4\over 3} R_{min}$.  For a young star with $L_* = L_\odot$, $T_* = 4000$\,K, and adopting $R_{min} = R_* \approx 2 R_\odot$, the wavelength reference is $\lambda(R_{min}) \approx 7\,\mu$m.  At the outer edge of the disk, we can define a similar ``reference'' wavelength, that, for $R_{max} = 100$\,AU, say, is $\lambda(100\,{\rm AU}) = 7$\,mm.  Emission from the disk dominates the SED throughout the far infrared, roughly $10 < \lambda < 100\,\mu$m.

\begin{figure}
 [!h]
  \begin{center}
    \leavevmode
  \centerline{\epsfig{file=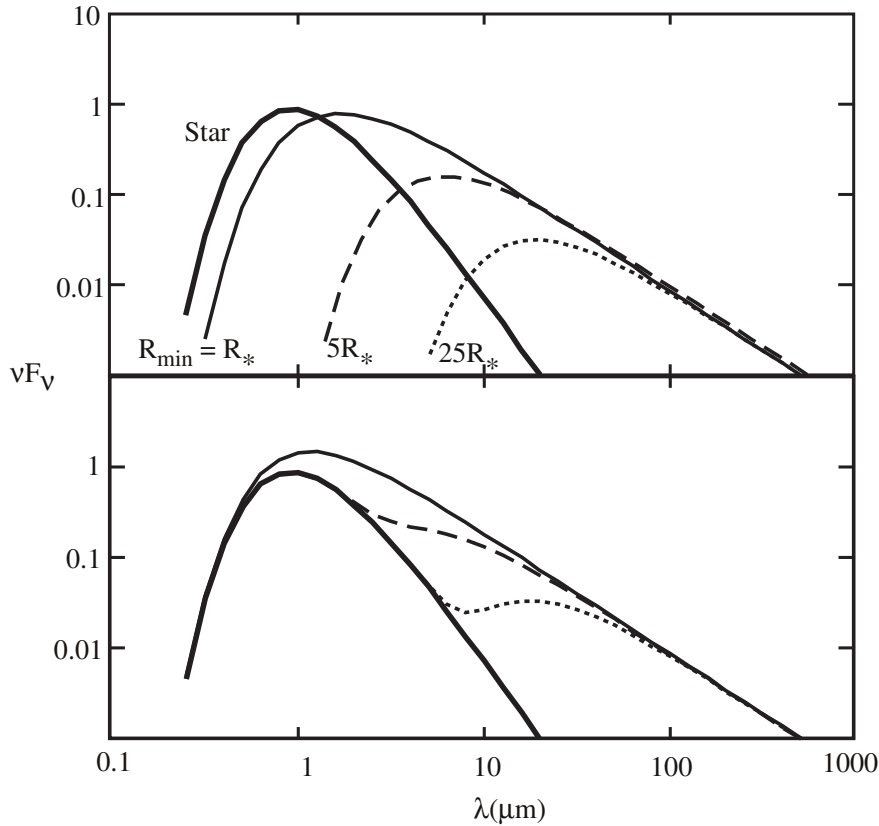,width=12cm}}
  \end{center}
  \vskip.1in
\caption{These spectral energy distributions were calculated from the equations for a flat disk with different assumptions about the inner disk radii as denoted in the figure. The top half shows the separate contributions to the SEDs from the star and the disks with various inner radii.  The bottom half shows the total SED consisting of the star alone, then the star plus the three disks with differing inner radii.}
\label{FlatSEDs}
\end{figure}

It is also interesting to see what happens as the inner radius is increased.  A disk with a hole in the center, such as seen around the main sequence stars Vega and $\beta$~Pictoris, will not radiate at short wavelengths, because there is not enough material near the star.  The effect is to lower the contribution from the disk in the near infrared without changing the far infrared radiation.  SEDs characteristic of disks with inner holes are seen, and the changes brought about by a hole are pretty much the same for complicated disk geometries with flaring and radiative transfer as they are for the flat, black disk.  Figure \ref{FlatSEDs} shows a number of SEDs for different assumptions about the flat disks developed in this section.  Keep in mind that some inner holes in the dust distribution will follow naturally from the melting of dust at $\sim 1600$\,K near the stars.

\begin{figure}
  \begin{center}
    \leavevmode
  \centerline{\epsfig{file=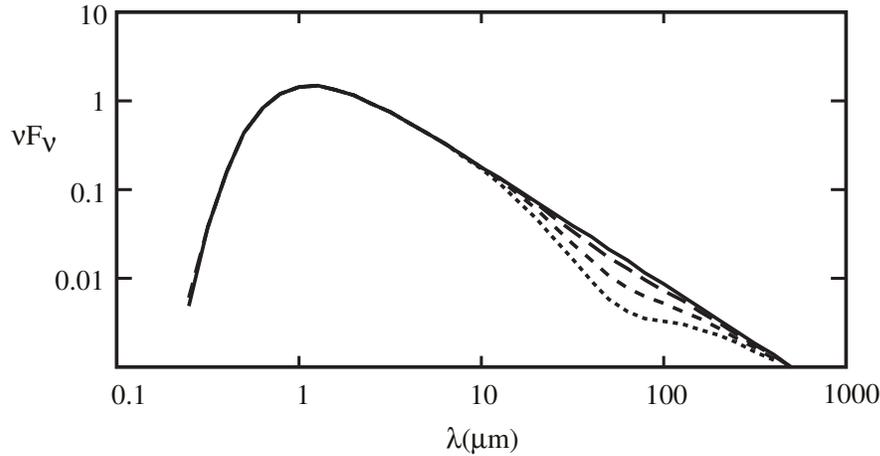,width=12cm}}
  \end{center}
  \vskip.1in
\caption{These spectral energy distributions were calculated assuming gaps of different sizes are created in an otherwise continuous disk.  The solid line is a continuous disk; the long dash is a disk with a gap between 0.75 and 1.25\,AU; the short dash is a disk with a gap between 0.5 to 2.0\,AU; the dotted line is a disk with a gap between 0.3 and 3\,AU. Notice that the effect of a gap is slight until the gap cuts out almost a decade of radii from the disk.}
\label{GapSEDs}
\end{figure}

If a gap in the disk develops, it will also change the SED.  Gaps are thought to be created when the first planets are assembled in disks.  Unfortunately, a gap will have only a modest effect on the SED, unless it is unusually large.  The reason for this is that every wavelength gets contributions from the disk over a fairly large range of radii, because of the width of the integrand in (\ref{FlatInteg}) and the slow variation of temperature as a function of radius.  Therefore, gaps will be difficult to detect from the SEDs alone, unless they are very large.  Figure~\ref{GapSEDs} gives several examples.

The flat, black disk model makes a number of predictions that can be compared with observations:
\begin{enumerate}
\item The infrared spectral energy distribution is a power law: $F_\nu \propto \nu^{1\over 3}$.
\item The total disk luminosity, as measured by emission in excess of that from the stellar photosphere alone, is about $1\over 4$ that of the stellar luminosity.  This fraction is exact for a disk extending from the stellar surface to infinity, and is in any case an excellent approximation for a disk extending a few AU from the star, since most of the luminosity comes from the inner radii.
\item The spectrum of the disk should be smooth, if it is truly black.  If there is some radiative transfer -- i.e. if the disk has finite thickness -- spectral features should appear in emission, because the disk atmosphere is heated from above.
\item If there is additional luminosity very close to the star, as may occur if radiation is released by accretion of matter onto the stellar surface, the disk emission will also increase in response to this additional heating.
\end{enumerate}

\begin{figure}
 [!h]
  \begin{center}
    \leavevmode
  \centerline{\epsfig{file=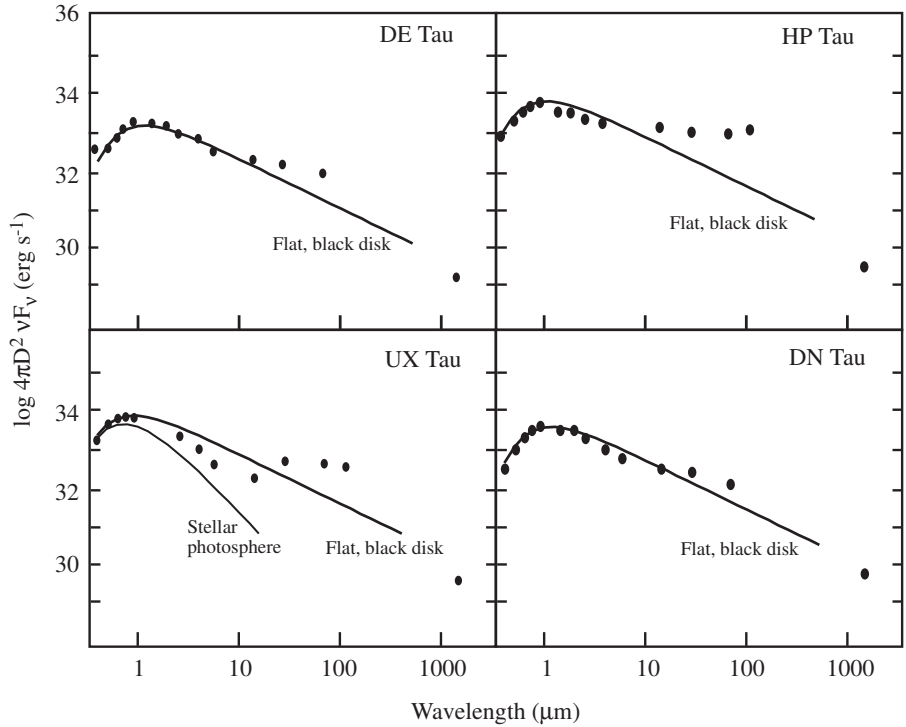,width=12cm}}
  \end{center}
  \vskip.1in
\caption{Observed spectral energy distributions are plotted on calculations of SEDs using the flat, black disk model.  In general, the actual SEDs have more excess infrared radiation than predicted by the flat disk model.}
\label{RealSEDs}
\end{figure}

It is easy to test these predictions.  Figure~\ref{RealSEDs} shows several observed spectral energy distributions of T~Tauri stars with excess infrared emission compared with the predictions of the flat, black disk calculations.  It is immediately obvious from the figure that the simple model does {\it not} fit the data.  T~Tauri SEDs generally have more emission at long wavelengths than the disk model permits.  The infrared luminosities are, therefore, higher, and the slope of the long wavelength SED is flatter than the model.  Modifications of the model by increasing the inner radius, $R_{min}$, or decreasing the outer boundary, $R_{max}$ relative to the ideal model would exacerbate the differences with the data.  The data show that if the T~Tauri stars are surrounded by disks, the outer regions of the disks are {\it hotter} than they would be if heated by the central sources in a flat disk model.  

This problem was recognized as soon as the first SEDs were observed (Rucinski 1985) and subsequently treated in an ad hoc manner by simply modifying the disk temperatures to empirically fit the SEDs (Adams, Lada, and Shu 1988).  The ad hoc assumptions about disk temperature are unsatisfying in absence of a physical theory.  The flat, black disk model is not adequate to fit the great majority of data on circumstellar disks.

The observations indicate that there is more far infrared radiation than expected from flat disks.  Either the outer radii are hotter than would be the case for flat disks, or there are additional components -- other reservoirs of dust -- that can absorb and radiate along with the disk. To heat the outer radii, one must either increase the radiation field at large distances from the star or modify the disk in some way to absorb more energy from the star.  There are models in the literature for all three approaches, additional dust components, increased radiation field, and increased absorption by the disk, and there is not yet a consensus on whether any or all of the possibilities are important in different objects.  

\subsection{Modifying the disk shape: flaring}
\begin{figure}
 [!h]
  \begin{center}
    \leavevmode
  \centerline{\epsfig{file=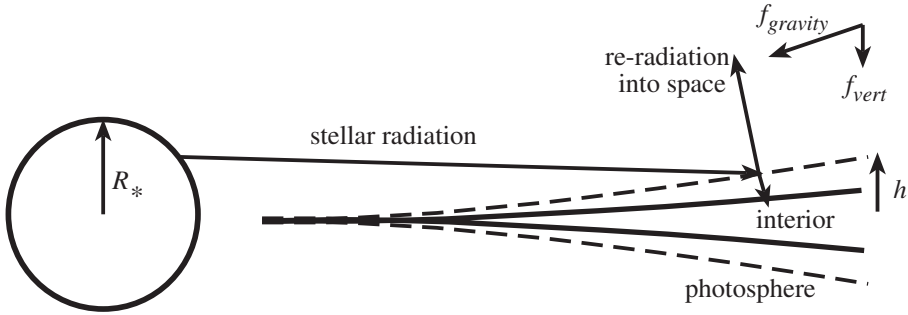,width=12cm}}
  \end{center}
  \vskip.1in
\caption{The flaring of a disk occurs naturally for a disk in hydrostatic equilibrium.  The disk mass is assumed to be negligible; gravity from the star acts to keep the material in a plane. The scale height of the disk increases with radius, because the thermal energy decreases more slowly than the vertical gravitational energy as radius increases. The vertical gravitational force, $f_{vert}$, is shown as a component of the stellar gravitational force, $f_{gravity}$. The ray from the star shows the point at which short wavelength stellar radiation from the star is absorbed in the disk photosphere. The two other rays from this point show how the energy is reradiated into space and into the interior of the disk, thus heating the interior from the above.}
\label{HydroFlare}
\end{figure}

Disks should not be completely flat. Flaring will occur naturally in the outer parts of a disk due to the warming of the material.  The disk should be in vertical hydrostatic equilibrium -- i.e. normal to the surface -- and ratio of thermal to gravitational energy increases with disk radius.  In the outer parts, thermal energy fattens the disk where gravity is too weak to confine the material to a thin plane.  For example, if the temperature as a function of radius, $T(r) \sim r^{-3/4}$, as for a flat disk, and the vertical gravitational potential is dominated by the central star, $E_{vert} \sim -{z\over r} {G M_* \over r} \sim k T(r)$, then the scale height will vary with radius as, $h_{scale} \sim {k \over G M_*} r^{5/4}$, and the disk will flare -- we ignore factors of order unity for this example.   Figure~\ref{HydroFlare} shows schematically what this means.

Flaring decreases the angle between the star and the normal to the surface, thus increasing the amount of light absorbed - the angle, $\phi$,in (\ref{FirstIllum}) is increased.  Kenyon and Hartmann (1987) recognized the importance of this effect as it applies to SEDs of disks.  They noted that many SEDs can be fitted well by adopting some modest flaring of the disk surface and modeled the flaring necessary to produce good fits to extant data with black disks.  Calvet et al. (1992) added the important effect of radiative transfer in a flared disk.  They realized that the disk atmosphere would have a temperature inversion if heated by the star, and they explicitly calculated models for flared disks with realistic atmospheres.  Their numerical calculations reproduced the main features of disk SEDs for a large number of disks and specifically predicted the presence of emission features, both molecular lines and resonant dust features, if the disks are irradiated from the central source.  

Chiang and Goldreich (1997) refined this idea by showing how the main physics, especially the flaring, could be understood analytically in a self-consistent calculation.  They showed that realistic models of disk atmospheres do a good job of fitting data for most but not all disks.  They also pointed out that because the dust particles that absorb the stellar radiation are in an optically thin part of the disk as far as re-emission is concerned and are small compared to the wavelengths of radiation, their equilibrium temperatures are warmer than they would be for purely black particles (cf Spitzer 1978).  The upper disk atmosphere is warmer at larger distances from the star than predicted by the black disk calculations.  Radiation at a specific wavelength must, therefore, have its origin over a larger area in the disk than expected from the black disk models.  It will be easier to resolve these regions when very high angular resolution becomes available than predicted by some estimates (e.g. Bertout, Reipurth, \& Malbet 1995).

The results of these calculations produce two main changes to our thinking about the physics of disks.  First, the disk flares approximately as a power law with radius, where the height of the visible photosphere above the midplane, $h_{phot}$, as a function of distance from the star, $r$, is: $h_{phot} \sim r^{58\over 45}$ (Chiang \& Goldreich 1997) or approximately $r^{4\over 3}$.  

Second, in contrast to the black disk where radiation is effectively absorbed and emitted from a black surface, light from the star at short wavelengths -- 0.5 to 2\,$\mu$m, say -- is absorbed in the visible photosphere at height, $h_{phot}$, where it heats the dust.  The dust then reradiates at longer wavelengths, where it is optically thin above this layer and also downward into the disk until it is absorbed by dust in a lower layer that is opaque to the longer wavelength radiation.  Because this longer wavelength radiation contains no more than half the energy originally absorbed in the photosphere, the inner layer is at a lower temperature than the photosphere.  The result is a disk atmosphere in which the temperature increases vertically from the midplane outward, and most of the radiation seen by the observer is optically thin.  Figure~\ref{HydroFlare} is a shows the geometry of the radiative transfer and the different layers.

\begin{figure}
  [!h]
  \begin{center}
    \leavevmode
  \centerline{\epsfig{file=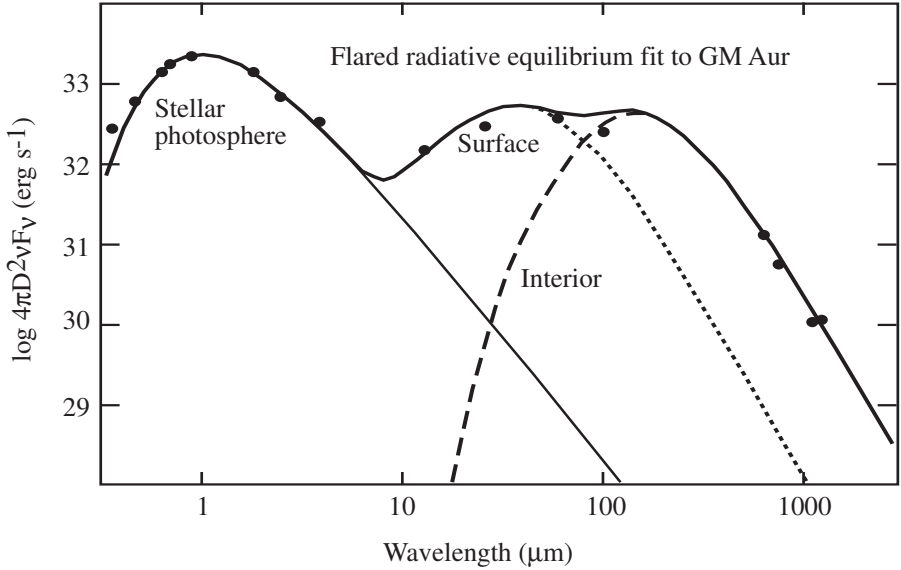,width=12cm}}
  \end{center}
  \vskip.1in
\caption{Figure~8 from Chiang and Goldreich (1997) showing how a flared disk with a photosphere reproduces one SED that differs substantially from a flat, black disk. They need a hole in the inner disk to account for the lack of disk emission short ward of about 5\,$\mu$m.}
\label{CGFig8}
\end{figure}

The combination of optically thin and optically thick emission from the photosphere and interior, respectively, creates an SED that flattens out or rises toward long wavelengths.  Figure~\ref{CGFig8} reproduces the calculation by Chiang and Goldreich (1997) showing how these models fit the data from the T~Tauri star, GM~Aur (see also Calvet et al. 1992).  The flared disk model can fit many of the disks in Figure~\ref{RealSEDs} with modest adjustments to the free parameters.  Since it is self-consistent and invokes no other components in the model except a star and a disk, it is a reasonable explanation for many observed SEDs.  The superheated, optically thin layer obeys the scaling, $F_\nu \sim \nu^{-5/7}$ ($q = {14\over 33}$), and the interior, optically thick layer scales as, $F_\nu \sim \nu^{-2/3}$ ($q = {6\over 15}$).

\begin{figure}
  \begin{center}
    \leavevmode
  \centerline{\epsfig{file=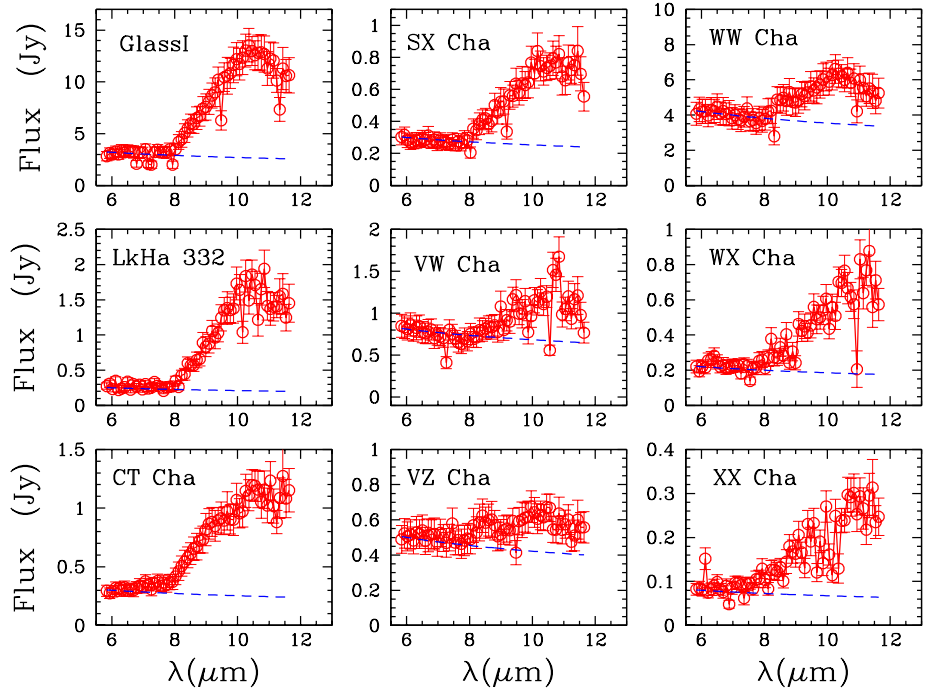,width=12cm,angle=-90,scale=0.8}}
  \end{center}
  \vskip.1in
  \caption{\em Observed spectra for the 9 stars in the sample from Robberto et al. (1999). The dotted line shows the continuum fitted to the data.  The emission features are due to a resonance feature in silicate dust around 10\,$\mu$m showing that these disks almost all have emission from optically thin dust, most likely in the superheated layers in the disk photospheres.}
\label{RFig5}
\end{figure}

\begin{figure}
  \begin{center}
    \leavevmode
  \centerline{\epsfig{file=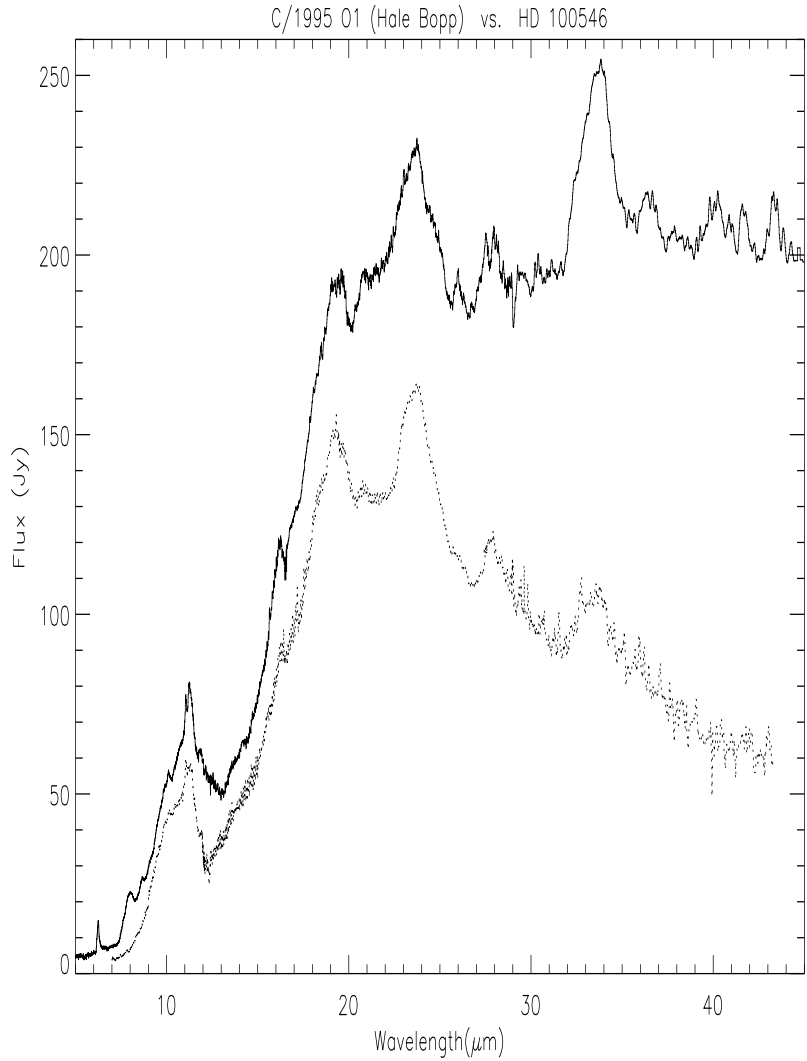,width=12cm}}
  \end{center}
  \vskip.1in
  \caption{\em This is Figure~5 from Malfait et~al. (1998) showing the detection of emission features in the disk around the Herbig Ae/Be star, HD~100546, (solid line at top) as predicted if the disk is heated by radiation from the central source. It is compared with the spectrum of comet Hale-Bopp (dotted line underneath).  Notice the close correspondence between emission features in the comet and in the disk spectrum, indicating that the particles in the disk are made from similar material as particles in the early solar system that made up the comet.}
\label{Waelkens}
\end{figure}

The prediction of Calvet et al. (1992) and Chiang \& Goldreich (1997) that irradiated disks produce emission features can be tested.  A strong emission feature near 10\,$\mu$m usually attributed to silicate resonance in dust was first observed for T~Tauri star disks by Cohen and Whiteborn (1985) and more recently by Robberto et al. (1999).  Figure~\ref{RFig5} reproduces Figure~5 from Robberto et al. (1999) showing spectra of 9 disks in the region of the silicate feature at 10\,$\mu$m, that can be compared with Figure~10 from Chiang and Goldreich (1997) showing the expected strength of emission features for a radiatively heated disk.  The most beautiful example of emission features is given by Waelkens et~al. (1998) and Malfait et~al. (1998) from their ISO spectra of the Herbig Ae/Be star HD~100546 compared with the spectrum of comet Hale-Bopp.  Figure~\ref{Waelkens} reproduces these spectra and demonstrates not only that emission features come from optically thin dust in the disk, but that these features indicate the dust composition is almost identical with that in the early solar system based on the comet spectrum.  The results support the conclusion that disks are radiatively heated from above, as in the flared disk model.

\subsection{Exceptions: Flat spectrum sources}
The disk SEDs that are most difficult to fit with the models described above are the flat spectrum sources (Adams, Lada, and Shu 1988).  The principal problem is that they emit more luminosity in the far infrared than at optical/near-infrared wavelengths, and so they defy explanations in which the disk temperatures are produced by radiation from the central star.  Since a classical accretion disk produces an SED of identical shape to that of a flat, black disk, $T(r) \sim r^{-3/4}$, it does not help to increase the accretion rate. What is needed is to increase the temperatures in the outer disk, add another component for the far infrared radiation, or decrease the optical radiation in a very selective way; the optical/near-infrared colors are usually inconsistent with pure extinction by normal interstellar dust.  All ideas to understand the flat spectrum sources include additional components to the star and disk.

A disk may be warmed by increasing the radiation flux on the outer parts if there is a way to scatter radiation onto the disk from the surrounding region.  Natta (1993) suggested that an optically thin spheroidal cloud, a ``halo'', might scatter starlight in the outer regions of the disk to increase the energy absorbed and, therefore, the temperature.  The spheroidal cloud may be a tenuous remnant of the molecular cloud from which the star and disk were created or it may be a transient part of the environment, such as residual infall or outflow in a wind.  The halo acts like a greenhouse, trapping radiation at large distances from the star where the total scattering optical depth approaches unity.

Calvet et al. (1994) expanded upon this theme to compute models of multi-component systems with a star, disk, outflowing wind, and a halo out of which the wind carves a cavity.  These calculations can successfully reproduce almost all observed SEDs in star/disk systems, but they include quite a few adjustable parameters, and they cannot be unique fits to the data.  Calvet et al. (1994) argue that the flat spectrum sources are really closer to Class I sources in the current nomenclature than the Class~II sources that concern us here, and so these models are perhaps less relevant to understanding disk physics than the ones described in the previous subsection.  Chiang \& Goldreich (1998) fit Class I sources using only a disk viewed edge-on.  They probably need an extra scattering component to account for the visible and near infrared light from some of these stars, but the number of assumption is, nevertheless, smaller.

\begin{figure}
  \begin{center}
    \leavevmode
  \centerline{\epsfig{file=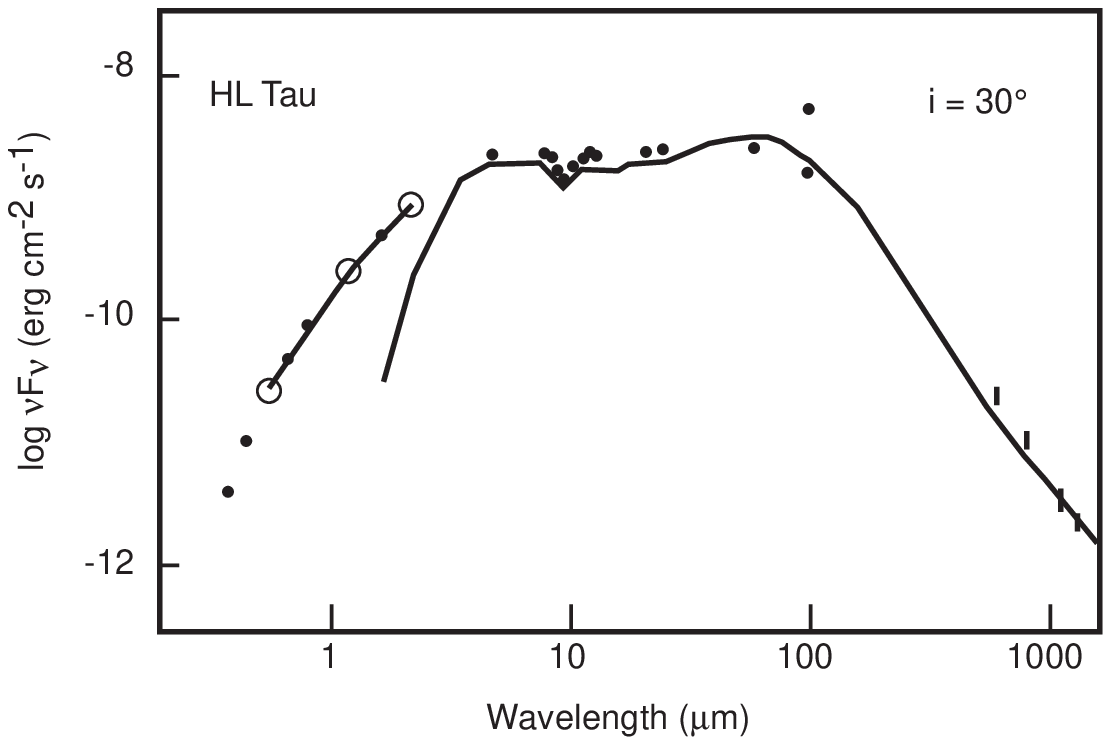,width=12cm}}
  \end{center}
  \vskip.1in
\caption{This is Figure~8 from Calvet et~al. (1994) showing their fit to the flat spectrum source, HL~Tau.  Their model includes a disk, an infalling envelope with a cavity along the polar axes with an opening angle of $10\deg$, and a scattering halo filling this cavity to produce the near infrared and optical light -- shown as a separate line connecting open circles at the shortest wavelengths.  This model successfully fits the SED at the expense of many adjustable parameters.}
\label{CalvetModel}
\end{figure}

Calvet's calculations demonstrate that realistic models of young stellar environments can fit the data in even the most truculent of SEDs.  The data do not require new physics or surprising physical conditions near the stars.  Figure~\ref{CalvetModel} is an example of Calvet's model applied to the flat spectrum source, HL~Tau.  Her models require many assumptions about the distribution of matter near the stars, including infalling envelopes with cavities, winds, and non-spherical density distributions.  In principle, most of these assumptions can be checked by observation, and it should be possible to determine how well the models correspond to reality.

The introduction of additional parameters into these models does call into question our ability to derive physical parameters from the limited data sets provided by SEDs.  Figure~\ref{RealSEDs} shows that the number of independent points in a typical SED is limited.  Often, there are at most five data points between 10\,$\mu$m and 2\,mm, the four IRAS measurements and a single millimeter-wave observation.  The theoretical treatment of flat disks shows that each point derives its emission from a wide range of radii in the disk, and that rather large changes in the disk -- large gaps, for example -- impress modest changes on the detailed SED, details that may often be confused with noise when only a few points are available.  The poor angular resolution of the IRAS observations means that far infrared data are susceptible to extended emission from the surrounding molecular cloud or a very tenuous but large halo surrounding the star.  It is challenging, perhaps impossible to sort out the various contributions to the SED in this case.  Therefore, the analysis of SEDs can give us only limited information about the disk upon which we can rely.  The most important way to disentangle the various components will need very high angular resolution in the infrared (e.g. Padgett et~al. 1999) and submillimeter.

\subsection{Very long wavelength emission: measuring disk mass}
At wavelengths longer than about 300\,$\mu$m, these disks become optically thin; you can see every particle in them.  Therefore, the spectral energy distributions at these wavelengths are linearly proportional to the total amount of dust, and it is possible in principle to figure out the disk mass from the observations.  The dust temperatures are almost always high enough to put this long wavelength emission on the Rayleigh-Jeans side of the Planck function, and the total emission becomes particularly simple to express.

The flux density, $F_\nu$, from an optically thin disk at distance, $D$, is:
\begin{equation}
F_\nu = {1\over D^2} \int_{R_{min}}^{R_{max}} B_\nu\left[T(r)\right] \tau_\nu(r) 2 \pi r dr, \label{low-tau}
\end{equation}
where $\tau_\nu(r)$ is the optical depth.  The optical depth can be written in terms of the surface density, $\Sigma(r)$, and mass opacity coefficient, $\kappa_\nu$ as: $\tau_\nu(r) = \kappa_\nu \Sigma(r)$.  If the emission is in the Rayleigh-Jeans regime, $B_\nu \approx 2 k T \nu^2 / c^2$.  In this limit:
\begin{eqnarray}
F_\nu & \approx & \kappa_\nu {2 k \nu^2 \over c^2 D^2} \int_{R_{min}}^{R_{max}} T(r) \Sigma(r) 2 \pi r dr \\
      & \approx & \kappa_\nu {2 k \langle T \rangle \nu^2 \over c^2 D^2} M_{dust}, 
\label{Fnu-long}
\end{eqnarray}
where $\langle T \rangle$, discussed explicitly in the next paragraph, is defined implicitly in (\ref{Fnu-long}).  Regardless of the disk structure, the flux density is directly proportional to the particle mass opacity, $\kappa_\nu$, and the total particle mass, $M_{dust}$, through the integral.  The absolute value of $\kappa_\nu$ is difficult to determine accurately, although a good estimate may be made with reference to interstellar clouds. The frequency dependence of $\kappa_\nu$ can be observed directly.

If the temperature were constant, the integral is just the mass of the disk times the temperature: 
\begin{equation}
\int_{R_{min}}^{R_{max}} \Sigma(r)\, 2 \pi r\, dr = M_{dust}.
\end{equation}
The variable temperature has the effect of weighting the hotter (inner) parts of the disk more than the cooler (outer) parts.  If the weighting function is not too steep, we can replace the integral in (\ref{Fnu-long}) with $\langle T \rangle M_{dust}$, where $\langle T \rangle$ is some suitable average temperature that takes care of the weighting functions.  We can calculate $\langle T \rangle$ explicitly for power law forms, $T(r) = T_0 (r/R_{min})^{-q}$, and $\Sigma(r) = \Sigma_0 (r/R_{min})^{-p}$:
\begin{eqnarray}
I & \equiv & \int_{R_{min}}^{R_{max}} T_0 \Sigma_0 \left( r \over R_{min} \right)^{-q - p} 2 \pi r dr \\
      & = & 2 \pi T_0 \Sigma_0 R_{min}^2 {\left( R_{max} \over R_{min}\right)^{2-q-p} - 1 \over {2 - q - p} } \\
   & = & T_0 M_{dust} {\left( R_{max} \over R_{min}\right)^{2-q-p} - 1 \over \left( R_{max} \over R_{min}\right)^{2-p} - 1 } { 2 - p \over 2 - q - p} \\
   & \approx & T_0 M_{dust} \left( R_{max} \over R_{min}\right)^{-q} {2 - p \over 2 - q - p},
\label{Tweight}
\end{eqnarray}
where we have assumed that $2 - q - p > 0$ for the last approximation and note that $R_{max}/R_{min} \gg 1$; typically, $R_{min} \sim 0.01\,$AU and $R_{max} \sim 100\,$AU  making this an excellent approximation.  If $2 - q - p = 0$ or $2 - p = 0$, one of the integrals is a logarithmic, and the approximation does not hold for $2 - q - p < 0$, but it is easy to work out the answers for those cases, too.  For the case in (\ref{Tweight}): 
\begin{equation}
\langle T \rangle \approx T_0 \left( R_{max} \over R_{min}\right)^{-q} {2 - p \over 2 - q - p}.
\end{equation}
Thus, for $q = 0.5$ and $p = 1$, $\langle T \rangle \approx 2 T(R_{max})$, and $F_\nu \approx \kappa_\nu {4 k T(R_{max}) \nu^2 \over D^2} M_{dust}$.  

Using the analytical results from Chiang and Goldreich's analysis for the interior disk layer that produces almost all of the long wavelength radiation, we can work out $\langle T \rangle$ for a centrally irradiated, flared disk of the sort in their model.  Noting that $q = 6/14$ and $p = 3/2$ in their calculations:
\begin{eqnarray}
\langle T \rangle & \approx & T_0 {\left(R_{max} \over R_{min}\right)^{1\over 14} - 1  \over \left(R_{max} \over R_{min}\right)^{1\over 2} - 1 }\ {{1\over 2} \over {1 \over 14}} \\
        & \approx & {1\over 2} T_0 \left( R_{max} \over R_{min}\right)^{-{1\over 2}} \ln\left( R_{max} \over R_{min} \right) \\
       & \approx & {1\over 2} T(R_{max}) \left( R_{max} \over R_{min} \right)^{1\over 14}  \ln \left( R_{max} \over R_{min} \right) \\
       & \approx & 2.4\, T(R_{max}).  \label{CGdustT}
\end{eqnarray}
For the numerical answer in (\ref{CGdustT}), I assumed that $(R_{max} / R_{min}) = 10^4$, but the very weak dependence of $\langle T \rangle$ on this ratio means that it is not a major uncertainty.  This result is useful as a simple way to estimate disk mass from the temperature in the outer regions and the observed flux density.  In most models including Chiang and Goldreich's, the temperature in the outer parts of the disk is around 20\,K.  This means that $\langle T \rangle \approx 50\,$K for the purposes of estimating disk mass.  In principle, it can be determined directly from millimeter-wave maps of molecules in the outer disk.  In any case, it is not very sensitive to the conditions in the disk for these models.

The major uncertainty in mass estimates is in $\kappa_\nu$, not in the model parameters.  Theoretically, $\kappa_\nu$ should be well determined for particles that are much smaller than the wavelength of observation, but it can vary substantially with changes in particle composition -- silicates vs. iron, for example -- and to some extent with particle size and shape.  Attempts to determine $\kappa_\nu$ from observations have given disparate results at millimeter wavelengths to date (Draine 1989; Beckwith \& Sargent 1991), with variations of an order of magnitude published for $\sim 1$\,mm wavelength.

Theoretically, $\kappa_\nu$ is expected to vary as a power of the frequency, $\nu$: $\kappa_\nu = \kappa_0 (\nu/\nu_0)^\beta = \kappa_0 (\lambda/\lambda_0)^{-\beta}$. A commonly adopted value at the time of this writing is $0.02 (1.3\,{\rm mm}/\lambda) \,{\rm cm}^2$\,g$^{-1}$ from Beckwith et al. (1990) and applies to total gas mass assuming $M_{gas}/M_{dust} = 100$, but this value is by no means certain and could be wrong by a factor of five in either direction.  Adopting this value, the disk mass can be written as a function of the continuum flux density at millimeter wavelengths as:
\begin{equation}
M_{disk} = 0.03\,M_\odot\ {F_\nu\over 1\,{\rm Jy}}\, \left(D\over 100\,{\rm pc}\right)^2\,\left(\lambda \over 1.3\,{\rm mm}\right)^3\,{50\,{\rm K} \over \langle T \rangle}\, {0.02\,{\rm cm}^2\,{\rm g}^{-1} \over \kappa_{1.3\,{\rm mm}}}.
\end{equation}

\subsection{Summary}
The spectral energy distributions of disks can reveal quite a bit of information about their physical properties.  Some of this information is robust, and some is model dependent and must be regarded with skepticism.  
\begin{itemize}
\item Disks have broad thermal emission from the near infrared to the submillimeter, much broader than a Planck function (blackbody).  The emission is often well fitted by a power law over an order of magnitude or more in wavelength.
\item The radial temperature dependence of the photospheric layer has an almost one-to-one relationship with the SED.  A good estimate of the temperature as a function of radius, $T(r)$, can be derived from the SED.
\item Typical disk temperatures are warmer than predicted by pure accretion heating or pure irradiation of a flat, black disk by a central source.  Flaring of the disks is an economical means of explaining the warmer temperatures for many but not all disks.
\item Large gaps, including inner holes, will create ``holes'' in the SED, as well, although the effect of gaps is diminished because each wavelength derives contributions from a wide range of radii in the disk.
\item Disks are expected to be optically thin (transparent) at millimeter wavelengths, and the total mass may be estimated from a measurement of the optical depth.
\item Disks with residual material in from their formation or from central outflows often display complex SEDs that require several components to reproduce.  The SEDs alone are inadequate to demonstrate that the model distributions reflect the actual disk environment, but the model predictions can be checked from observations with high enough angular resolution.
\end{itemize}

\section{Properties of Disks}
It is not possible to image most disks in the manner of Figure~\ref{DiskImages}.  They are rarely situated in front of bright background sources, such as those in the top half of the figure, and they are normally not sufficiently edge-on to be seen in scattered light as they are in the bottom half of the figure.  Detecting disks for the purposes of counting is normally done by looking for infrared radiation in excess of the stellar photosphere and assuming that any excess is due to disks.

\subsection{How many young stars have disks?}

As a general rule, it is easier to infer the presence of disks from radiation at wavelengths long ward of 10\,$\mu$m, say, because that radiation is generated far from the star and requires more than trace amounts of dust to be observable. The first attempts to count disks in samples of young stars used far infrared and millimeter wave observations.  Strom et al. (1989) and Cohen, Emerson, \& Beichman (1989) found that nearly half of the young stars in their samples were detected in one or more of the IRAS bands -- 12, 25, 60, and 100\,$\mu$m -- and were likely to have disks.  Beckwith et al. (1990) looked at similar samples at 1.3\,mm detecting about 40\% of them.  These early studies asserted that disks were common among classical T Tauri stars.

A variety of different surveys since then have generally supported the finding that disks are common, although the exact frequency depends on the sample and the selection criteria for young stars.  It is much easier to detect light in the near infrared, at 2\,$\mu$m, for example, where large arrays of sensitive detectors are readily available, but the amount of radiation from a disk is a much smaller fraction of the total -- the star is bright, too -- and very little dust is needed to produce an excess.  Surveys of dark clouds at 2\,$\mu$m show disk fractions in anywhere from 10\% to 90\% of the sample.  The high figure, 90\%, is seen in the innermost regions of the Orion Trapezium cluster (McCaughrean and Stauffer 1994), whereas the low figure includes large numbers of stars selected on the basis of x-ray emission (Walter et al. 1988).  It is still controversial if x-ray emission is sufficient by itself to reveal very young stars, so the 10\% figure should be regarded with caution.  Recent surveys at 15\,$\mu$m with the ISO satellite give numbers in the 30 to 50\% range (Nordh et~al. 1996; Nordh et~al. 1998; Bontemps et~al. 1998). It is reasonable to conclude that disks accompany young stars about half the time, and it may well be that all stars are born with disks.

\subsection{Disk lifetimes}
The lifetime of the dust is determined by counting disk fractions in young clusters with different ages.  The first systematic attempts were made by Beckwith et al. (1990) at 1.3\,mm and Skrutskie et al. (1990) at 10\,$\mu$m.  The latter paper found that the sample fraction with disks decreased markedly between ages of a few million and ten million years.  More recent studies using near-infrared excesses shows a steady decrease to almost zero at 10 million years (E. Lada; this volume) indicating that the inner parts of the disks disappear over this time.  A similar study at 25 and 60\,$\mu$m shows the outer parts of the disks disappear just as rapidly (Robberto et al. 1999; Meyer et al. 1999).  

Main sequence stars can also have tenuous disks, although the disk masses are many orders of magnitude less than their young counterparts (Backman \& Paresce 1993).  These disks are presumably the remnants of those in the young stars, and they may be generated by the grinding of larger bodies -- an asteroid belt, for example -- orbiting the stars.  Zuckerman and Becklin (1993) inferred a decrease of disk mass with time following approximately: $M_{disk} \sim t^{-2}$ during the first 300\,Myr.  This result suggests a steady loss of small particles over almost the entire pre-main sequence lifetime of a star.

Small particles should coagulate into larger ones on even faster timescales.  The early solar nebula is thought to have created small planets within 10\,Myr.  The observed disk lifetimes are, therefore, consistent with expectations for disks making planets, although there is no strong evidence that the particles are being accumulated into planet-sized bodies.

\subsection{Disk masses}
Most of the disk mass should be gaseous hydrogen and helium.  The particles probably contain only 1\% of the mass when a disk is first assembled from interstellar clouds.  This ratio may change as the disk evolves.

\begin{figure}
 [!h]
  \begin{center}
    \leavevmode
  \centerline{\epsfig{file=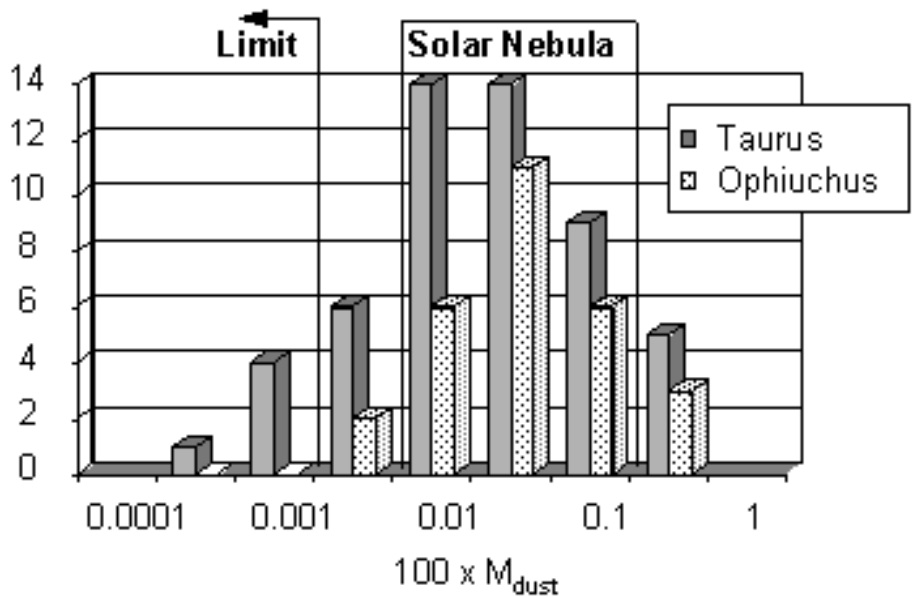,width=12cm}}
  \end{center}
\caption{The histograms show the distribution of disk masses among stars in the Taurus and Ophiuchus star forming regions determined by Beckwith et al. (1990)and Andr\'e et al. (1994).}
\label{MassDistribution}
\end{figure}

The particles are easy to observe, and the total particle mass can be estimated from the observed optical depth at millimeter wavelengths, as we saw in a previous section.  Millimeter wavelength observations of large samples show that typical disks are a few percent of a solar mass with a wide range.  Figure~\ref{MassDistribution} shows a histogram of disk masses for samples in the Taurus and Ophiuchus dark clouds (Beckwith et al. 1990; Andr\'e et al. 1994; Osterloh and Beckwith 1995).  These samples yield consistent results.  New data (E. Lada 1999, this volume) indicate that the disks in the Orion star forming region may have lower masses on average, although the sensitivities of the observations are not yet good enough to make a robust conclusion.

Theories of the early solar system require between 0.01 and 0.1\,M$_\odot$ in the solar disk to create our planetary system.  Disk masses around other stars fall in this range.  It is clear that there is enough mass in a typical disk to create a planetary system like our own, but there is no proof that it will do so. 

If disks do make planets, there should be evidence of particle growth in the older disks.  We know that particles are lost as the disks age.  If the loss mechanism is accumulation into larger bodies, it seems likely that planetary systems are the final product of this evolution; if there is no particle growth, the disks may simply disperse into space or be accreted onto the central stars.

\subsection{Particle sizes}
At submillimeter wavelengths and longer, $\kappa_\nu$ is expected to vary as a power of the frequency or wavelength: $\kappa_\nu \propto \nu^\beta \propto \lambda^{-\beta}$.  For compact spherical particles smaller than the observing wavelength, $\beta = 2$ for metals and insulators under a wide range of conditions (Bohren and Huffman 1983; Emerson 1988).  The power law exponent, $\beta$, can be as small as 1 for certain types of materials - amorphous carbonaceous material, for example - and even smaller over limited ranges, but it must go to 2 at long enough wavelengths to fulfill causality relations. It is expected that $\beta \approx 2$ for interstellar dust particles at long wavelengths.  Careful attempts to measure both $\kappa_\nu$ and $\beta$ in interstellar clouds have, indeed, yielded $\beta \approx 2$ and values of $\kappa_\nu =(0.002--0.004)(\lambda /1.3\,{\rm mm})^{-2}$\,cm$^2$\,g$^{-1}$ (Hildebrand 1983; Draine and Lee 1984).  If a mixture of particle types and sizes is present, the observed value is an average over the different constituents.  

Rocks, asteroids and planets are opaque to radiation at $\lambda \sim 1$\,mm, in which case $\beta = 0$.  Pebbles with sizes of order 1\,mm should have exponents with intermediate values: $0 < \beta < 2$.  If dust grains grow large enough to put most of the mass in bodies larger than pebbles, they should have an observable signature at these wavelengths.  Because the radiation does not penetrate far below the surface of a large body, meaning most of the material never interacts with it, the absolute value of the opacity per unit mass must decrease as the particles grow.

\begin{figure}
  \begin{center}
    \leavevmode
  \centerline{\epsfig{file=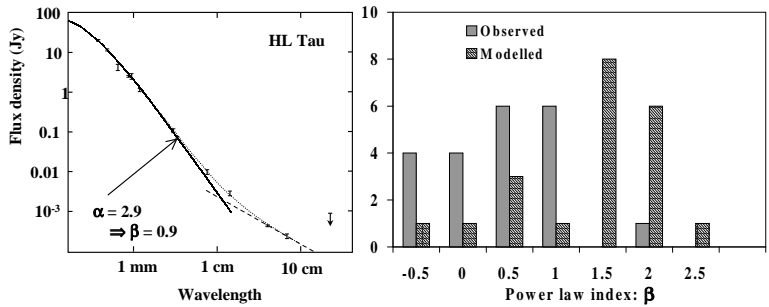,width=12cm}}
  \end{center}
\caption{The left hand part shows the spectral energy distribution of HL Tau from about 350\,$\mu$m to 6\,cm wavelength. The solid line fits the short wavelengths -- thermal emission by dust -- and the dashed line fits the long wavelengths -- free-free emission from ionized gas.  The long wavelength spectral index of the dust is 2.9, meaning $\beta = 0.9$ with no corrections for optically thick emission (Wilner, Ho, \& Rodriguez 1996).  The right hand portion shows the distribution of $\beta$'s derived in two ways for a larger sample of disks.  The ''observed'' $\beta$'s are derived directly from the spectral index as described in the text: $\beta_{obs} = \alpha - 2$.  The ''model'' $\beta$'s are derived by assuming a surface density distribution as discussed in the text (Beckwith \& Sargent 1991).}
\label{beta-distribute}
\end{figure}

Changes in the spectral energy distributions (SEDs) of disks relative to those typical of interstellar clouds are observed (Beckwith and Sargent 1991; Mannings and Emerson 1994; Koerner et al. 1995).  The flux densities, $F_\nu$, of disks tend to fall more slowly at wavelengths longer than about 400\,$\mu$m than those of interstellar clouds.  The spectral index, $\alpha$ - where $F_\nu \propto \nu^\alpha$ - is typically between 2 and 3, whereas $\alpha \sim 4$ for the interstellar medium.  Assuming optically thin emission in the Rayleigh-Jeans limit, the exponent, $\beta$, is between 0 and 1 for disks compared to the theoretical value of 2 for small particles in the long wavelength limit.  The right hand part of Figure~\ref{beta-distribute} shows the distribution of emissivity exponents, $\beta$, determined from the observed spectral slope of disk emission near $\lambda \sim 1$\,mm.  Many of the disks in this sample have $\beta \le 1$.  It is tempting to interpret this change as the result of particle growth.  

Figure~\ref{beta-distribute} includes the distribution of $\beta$'s derived from model fits assuming the surface density is proportional to $r^{-3/2}$.  Including the surface density distribution tends to increase the derived value of the emissivity exponent relative to the observed spectral slope.  For example at 1.3\,mm, observations of a disk with $M_D = 0.03\,$M$_\odot$, $R_D = 100\,$AU, and $\kappa_\nu = 0.02\,$cm$^2$\,g$^{-1}$, the average optical depth, $\bar\tau_\nu = 0.24$, and $\beta \approx 1.36(\alpha - 2)$.  Any conclusions about particle growth depend on knowledge of how much optically thick parts of the disks contribute to the spectral energy distribution at long wavelengths.  It is probably for this reason that the samples of disks with measured $\alpha$'s at long wavelengths have not been enlarged much since the work of Beckwith \& Sargent (1991) and Mannings \& Emerson (1994).

To measure the optical depth, it is essential that the disk emission be resolved at wavelengths where it is relatively transparent so that the surface density distribution may be determined directly from the distribution of optical depth.  Resolving the disks is time-consuming with the present suite of millimeter facilities.  Several groups have now managed to resolve the emission from one star, HL~Tau, by carrying out novel experiments with existing millimeter-wave telescopes.  Lay et al. (1994) and Lay, Carlstrom \& Hills (1997) combined the CSO and JCMT into a single-baseline interferometer to observe at 0.65 and 0.87\,mm and combined it with data from the Owens Valley millimeter interferometer at 1.3\,mm.  Mundy et al. (1996) extended the BIMA array to measure a size of the disk at 2.7\,mm wavelength, and  Wilner, Ho \& Rodriguez (1996) used the VLA to get sub-arcsecond resolution of the disk at 7\,mm.  Each group resolved the emission but with beam sizes of the same order as the disk size, and the sub-millimeter observations had $u-v$ plane coverage too poor to map the brightness distribution.  All groups measured the average brightness temperature and overall disk orientation and constrained subsequent model fitting much more than is possible from the SEDs alone.

\section{Are Disks Important?}
Disks are common around young stars, and they contain enough mass to make planetary systems like our own.  At the time of this writing, sixteen planets orbiting stars other than the Sun have been inferred using radial velocity observations of the stars -- although there is some controversy as to whether each of the objects is a planet according to usual definitions.  Since we know some extrasolar planets exist and there are many disks, it is a small leap to suppose that disks make planets, as the theory asserts, and there are many planetary systems.  This is an assumption, however, and proof is not likely to be forthcoming until we have actually observed some planets built up within the disks.

If planetary systems are common and disks are the necessary precursors, disks are important for the understanding of planetary systems.  They are likely to be our only means of filling in many missing links in our understanding of early Solar System evolution.  To understand our own origins and to assess the likelihood of other planetary systems, further study of disks around other stars presents one of the most fruitful avenues.

On the other hand, the disks do not appear to be of great importance to the early evolution of the star after the main collapse has occurred.  From the T~Tauri stage until the main sequence, disks probably contain only a few percent of a stellar mass, and the amount of material accreted onto the star during this period is not likely to be sufficient to add significantly to a star's mass.  It is possible that magnetic coupling between the disk and the star governs stellar rotation and determines the star's final rotation rate. There are some hints of this process but no proof.

Many of us suspect that the very early stages of disk formation are crucial to the buildup of a star.  During the initial collapse, there is almost certainly too much angular momentum in the molecular cloud to make a centrally condensed star unless a mechanism exists to shed angular momentum during the collapse.  A disk is a natural repository of such angular momentum.  In this way, disks may be essential to a star's birth.  But the disk angular momentum appears to be insufficient to take up that from the molecular cloud by the T~Tauri phase (Beckwith et~al. 1990), suggesting a large redistribution of angular momentum within the disk itself subsequent to the collapse phase.  It is still difficult to study disks in great detail during and immediately after the collapse phase.  There are not many candidates, yet -- the entire phase is very short-lived -- and the extinction toward the very young disks is too high to allow even near infrared observations.  Very high resolution millimeter wave images enabled by planned large interferometers (MMA/LSA) will likely change this situation and make observations of the physics of young disks routine.

\section*{ACKNOWLEDGMENTS}
I am grateful to Charlie Lada and Nick Kylafos for their generosity in inviting me to the Crete conference and to the postdocs and students of the conference for their stimulating questions and comments on circumstellar disks.  Portions of the text were taken from articles that I co-authored with Anneila Sargent, Thomas Henning, and Yoshitsugu Nakagawa; I am grateful for their permission to reprint those parts in this article.

\end{document}